\documentclass[twocolumn,twoside,slac]{revtex4}
\usepackage{graphicx}
\usepackage{fancyhdr}
\begin{document}

\title{Relational databases for data management in PHENIX}

\author{I.Sourikova, D.Morrison for the PHENIX collaboration}
\affiliation{BNL, Upton, NY 11973, USA}

\begin{abstract}
  PHENIX is one of the two large experiments at the Relativistic Heavy
  Ion Collider (RHIC) at Brookhaven National Laboratory (BNL) and
  archives roughly 100TB of experimental data per year.  In addition,
  large volumes of simulated data are produced at multiple off-site
  computing centers.  For any file catalog to play a central role in
  data management it has to face problems associated with the need for
  distributed access and updates.  To be used effectively by the
  hundreds of PHENIX collaborators in 12 countries the catalog must
  satisfy the following requirements: 1) contain up-to-date data, 2)
  provide fast and reliable access to the data, 3) have write
  permissions for the sites that store portions of data.  We present
  an analysis of several available Relational Database Management
  Systems (RDBMS) to support a catalog meeting the above requirements
  and discuss the PHENIX experience with building and using the
  distributed file catalog.

\end{abstract}

\maketitle

\thispagestyle{fancy}

\section{Introduction}
\begin{figure}[t]
\centering
\includegraphics[height=60mm,width=80mm]{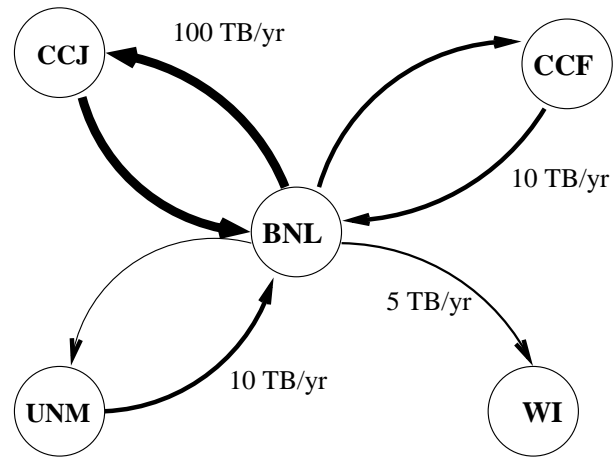}
\caption{Subset of PHENIX data transfers.
 CCJ - PHENIX Computing
Center in Japan, imports data for analysis, exports reconstructed and
simulated data. CCF - PHENIX Computing Center in France, large
facility similar to CCJ. UNM - University of New Mexico, imports data for analysis, exports simulated
data. WI - Weizmann
Institute (Israel), imports data for analysis.} \label{datatrans-f1}
\end{figure}

PHENIX began data taking in 2000 and has accumulated more than 400,000
data files.  In addition to data produced by the detector itself,
significant amounts of reconstructed and simulated data are produced
at PHENIX computing centers in the US, Japan and France.  These data
are then transferred to the central repository at BNL and from BNL to
off-site institutions for analysis.  Fig.~\ref{datatrans-f1} shows a subset
of PHENIX data transfers.

The information about files and their multiple replicas are kept in
the file catalog.  During the first year of data taking when most of
the data processing was done at the BNL RHIC Computing Facility (RCF),
a master copy of the file catalog was updated at BNL and replicated in
master-slave mode to remote sites.  This replication mode provided
remote sites with read-only copy of the file catalog.

When several PHENIX sites started large-scale production, this
read-only access proved--not surprisingly--to be unworkable.  Updating
the central catalog over the wide-area network was prohibitively slow
and the catalog was updated offline after the production.  This
required a lot of manual work and file catalog updates tended to lag
unacceptably far behind the production curve.
 
One solution to the problem would be to provide remote sites with
write access to a local file catalog database.  We started to look for
technology supporting peer-to-peer (also known as ``update anywhere''
or ``multimaster'') database updates.

\section{Objectivity peer-to-peer replication}

\begin{figure}[t]
\centering
\includegraphics[height=40mm,width=80mm]{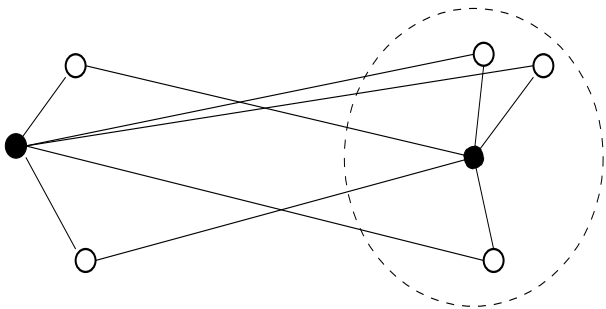}
\caption{Topology for Objectivity replication} \label{ObjyTopology-f2}
\end{figure}

PHENIX has an existing Objectivity-based file catalog, and although
Objectivity supports peer-to-peer replication, it requires that all
database clients have access to all servers that are involved in the
replication.  When one of the servers has a surrounding firewall (the
case of the file catalog at BNL), opening the requisite firewall
conduits for all outside clients raises security issues and exacts a
high maintenance price. Fig.~\ref{ObjyTopology-f2} illustrates the topology in
the case of two servers.  Black circles represent database servers and
white circles - database clients.  A dashed circle around one server
represents a firewall.  We realized that to provide production sites
with the ability to write to the local database another database
technology was needed.

\section{Database technology choice}
An obvious candidate for a database with support for peer-to-peer
replication was Oracle, but cost and licensing concerns led us to
focus instead on freely available solutions.  We considered three
open-source databases: SAP DB ~\cite{sapdb}, MySQL ~\cite{mysql} and
PostgreSQL (also known as Postgres) ~\cite{pg}.  At the time we
evaluated our technical options, neither SAP DB nor MySQL supported
peer-to-peer replication.  And although Postgres does not currently
have built-in support for replication, we found two active,
third-party projects providing Postgres peer-to-peer replication:
PostgreSQL Replicator ~\cite{pgreplicator} and Postgres-R ~\cite{pgR}.

Among the attractive Postgres DBMS features were the following:
\begin{itemize}
\item Postgres is ACID compliant, meaning that it is guaranteed to be
 in a consistent state at all times.
\item It has ``Multi Version Concurrency Control'' which enables it to
  scale well with a large number of concurrent applications
\item It adheres closely to the SQL'92 standard and has extensive
  documentaion, making it particularly easy to use
\item It has a sophisticated implementation of triggers, which allow
  particular actions to be initiated by activity involving the database
\item It has LISTEN and NOTIFY support message passing and client
  notification of an event in the database.  This can be used to
  automate data replication.
\end{itemize}

\section{PostgreSQL Replicator}
To build a distributed Postgres database with peer-to-peer updates we
used PostgreSQL Replicator (PR) software~\cite{Cavalleri} which
implements asynchronous (store and forward) Postgres replication.  PR
stores operations performed on a database in a local queue for later
distribution by a database synchronization process.

PR has a table level data ownership model which include Master/Slave,
Update Anywhere and Workload Partitioning table ownerships.  PR also
has a convenient table level replication that enables one to replicate
specific parts of a database.

\section{Distributed PHENIX file catalog}

We have implemented a distributed file catalog that currently includes
three sites: BNL, Stony Brook University (SBU) and Vanderbilt
University (VU). Each site runs a local copy of Postgres-based file
Catalog that has information about files on all three sites.

\subsection{Content of the file catalog}
All officially produced PHENIX files have a master copy in HPSS at RCF
and possibly one or more replicas on disk at various PHENIX
institutions.  The file catalog contains the following information:
\begin{itemize}
  
\item Logical file names: each officially produced file has a unique
  name - logical file name, which serves as a file identifier and
  provides for easy file relocation.
  
\item Hosts: physical machines with local data storage. A host can be
  a computer with local disks or HPSS with tape storage.

\item Clusters: a cluster is a set of NFS-interconnected hosts. 

\item Sites: PHENIX institutions that have one or more clusters.

\item Storage: disk or tape.
  
\item Link cost: a relative cost to transfer the file from site A to
  site B. Depends on the distance between A and B, network
  connectivity and possibly on some policies.

\end{itemize}

More detailed metadata about the physics content of each file is
maintained in a central Objectivity Run/File database.  To find a set
of files suitable for a particular physics analysis, users would
consult the Objectivity once and then use the file catalog database to
find the frequently changing information about physical locations of
those interesting files.

The file catalog can satisfy the queries like
\begin{itemize}
\item find all replicas of the file
\item find all disk replicas of the file
\item find a local disk replica of the file
\item list all files at site A
\item show all machines in cluster B
\item list all the files from a particular production
\item find 'closest' replica of the file
\end{itemize}

where 'closest' means the smallest link cost.

\subsection{File catalog replication}

Database tables that store the information about files and their
locations are replicated among all sites.  The replicated tables have
Workload Partitioning data ownership which means each site has
read-only access to the replicated partition of a table that comes
from another site and read-write access to local entries of the table.
In this data ownership model no conflicts can arise - each site
manages their own entries but is able to read the entries made by
other sites.

Each site that runs a local file catalog copy is responsible for
cataloging local files.  For example, at BNL the file catalog is kept
current by production jobs, by jobs that stage files from Mass Storage
System (HPSS) to disk and cron jobs that delete entries for the files
that have been deleted from disk.

\begin{figure}[t]
\centering
\includegraphics[height=70mm,width=80mm]{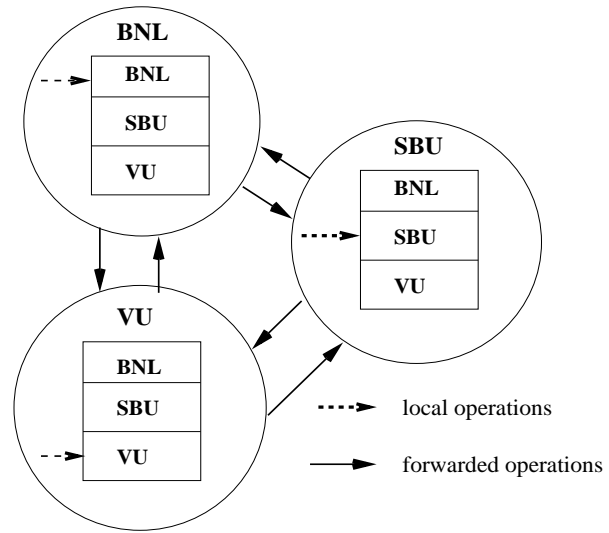}
\caption{Distributed PHENIX file catalog. BNL - Brookhaven National
Laboratory, SBU - Stony Brook University, VU - Vanderbilt
University.} \label{3sites-f3}
\end{figure}

Updates to local instances of the file catalog are propagated
periodically by database synchronization processes that restore the
data integrity of local catalogs.  This is shown schematically in
fig~\ref{3sites-f3}. Each replicated database table consists of three
partitions (which is equal to the number of sites) and only local
partitions accepts updates. Although the replication knows about data
partitioning, the data is still in one database table, which makes
querying the Catalog extremely easy. For example to find all replicas
of a file 'XYZ', one issues a query ``{\tt SELECT * FROM files WHERE
  lfn='XYZ'}``.

If the synchronization processes fails because one of the servers is
down or unreachable, it is repeated by a cron job until it succeeds.
In the meantime all the sites that have their local databases up and
running, can continue to update their file catalogs. That eliminates a
potential single point of failure of the central file catalog.

\subsection{Scalability issues}  

An important feature of PR is that replication is incremental and only
updates that were committed to the database since last synchronization
are transferred over the network.  This replication feature provides
an excellent scalability with database size and allows to maintain a
very high-level of synchronization between sites involved in the
replication.  The replication of 20,000 new updates between 3 sites
takes about 1 minute.  It depends on network ``weather'' and the
number of sites.

After synchronization, each local databases contains a view of the
entire data set.  Therefore, the local clients do not need to access
the remote database server.  That not allows us to limit the number of
firewall conduits to just server-server connections.

\section{Modifications to improve database replication scalability}

Although PR replication is incremental, when a new site is added to
the distribution, the entire database must be replicated to upload all
the existing data to the new site.  Since the replicated database
tables remain unwritable during the synchronization process (to
guarantee data consistency) adding a new site can cost several hours
of interrupted production.  To make the production able to update the
file catalog database independent of database replication, we have
added small modifications to the replication schema.
\begin{figure}[b]
\centering
\includegraphics[height=20mm,width=80mm]{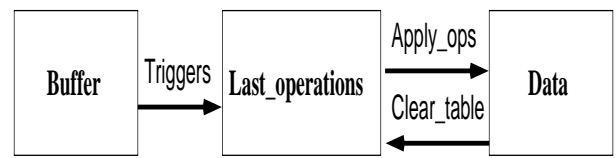}
\caption{modified replication schema} \label{staging-f4}
\end{figure}
  
The new schema is depicted in Fig.~\ref{staging-f4}.  The main idea is
to use deferred database updates.  All database operations are applied
to a buffer and then propagated to the replicated table.  Database
triggers capture the operations applied to the buffer table and write
them to the Last\_operations table.  After a cron job applies the
operations from the Last\_operations table to the Data table, the
Buffer and the Data tables becomes identical and the Last\_operations
table is cleared. The Data table is replicated periodically to remote
sites and during the replication the file catalog database accepts new
updates to the Buffer.

Our modification introduces a small (few minutes at most), acceptable
delay in seeing new files after they were put into the database, but
solves the problem of the write locks during adding a new site to the
distributed file catalog.

\section{File catalog API}

Although we need the file catalog to address the challenge of not
having enough resources to keep an entire data set on disk at all
participating institutions, we are trying to shield the users doing
physics analysis from the need to know where the files interesting for
their analysis are.

A file catalog C++ API provides a DBMS-independent layer that shields
the users from the file catalog implementation and performs a
logical-to-physical file name translation.  The users ask for the
files by logical file names and the analysis code need not be changed
when the files are relocated.  The file search is controlled by an
environment variable that can contain a colon separeted list of local
directories and/or reserved words for the database search.

In addition to C++ API, both perl and Web interfaces have been created
to update and query the file catalog.

\section{Summary}
Using Postgres DBMS and Postgres Replicator we have built a
distributed file catalog with multimaster updates that satisfies all
the initial requiremnts:
\begin{itemize}
\item Contain up-to-date data
\item Provide fast and reliable access to the data
\item Have write permissions for the sites that store portions of data
\end{itemize}

We introduced deferred database updates to the Postgres Replicator
schema to overcome the problem of database write locks during addition
of new sites to the production file catalog database.  We plan to add
three new production sites to the distibution.

\begin{acknowledgments}
  The authors wish to thank A. Shevel, I. Ojha and E. Vaandering for
  their work on distributing PHENIX file catalog database.

  Partial support for this work was provided by the National Science
Foundation grant PHY-0219210.

\end{acknowledgments}


\begin{thebibliography}{9}  

\bibitem{Cavalleri}
M.Cavalleri,R.Prudentino,U.Pozzoli,G.Reni ``A set of tools for
building PostgreSQL databases in biomedical environment'', 22th Annual
International Conference of the IEEE Engineering in Medicine and
Biology Society, Jul 2000
\bibitem{sapdb}
http://www.sapdb.org/
\bibitem{mysql}
http://www.mysql.com/
\bibitem{pg}
http://www.postgresql.org/
\bibitem{pgreplicator}
http://pgreplicator.sourceforge.net/
\bibitem{pgR}
http://gborg.postgresql.org/project/pgreplication

\end{thebibliography}
\end{document}